\documentclass[prl,aps,amsfonts,amssymb,nofootinbib,amsmath,superscriptaddress]{revtex4-2}

\usepackage{graphicx}
\usepackage{bm}
\usepackage{hyperref}
\usepackage{float}
\usepackage{siunitx}
\usepackage{xcolor}
\usepackage{natbib}
\usepackage[utf8]{inputenc}

\begin{document}

\title{Much ado about MOFs:  Metal-Organic-Frameworks as
Quantum Materials }

\author{N. Drichko} \email{drichko@jhu.edu}
\affiliation{William H. Miller III Department of Physics and Astronomy, The Johns Hopkins University, Baltimore, Maryland 21218, USA}

\author{V. S. Thoi}\email{sarathoi@jhu.edu}
\affiliation{Department of Chemistry, The Johns Hopkins University, Baltimore, Maryland 21218, USA}
\affiliation{Department of Materials Science and Engineering, The Johns Hopkins University, Baltimore, Maryland 21218, USA}

\author{N. P. Armitage} \email{npa@jhu.edu}
\affiliation{William H. Miller III Department of Physics and Astronomy, The Johns Hopkins University, Baltimore, Maryland 21218, USA}

\begin{abstract}

Metal-organic frameworks (MOFs) are a highly tunable class of crystalline materials where metal atoms or clusters are connected by organic linkers.  They offer a versatile platform for exploring quantum phenomena such as novel magnetism, superconductivity, and topology.  Particularly for magnetism, their modular chemistry enables extensive control over  interactions, spin magnitudes, lattice geometries, and even light-responsiveness, making them uniquely adaptable platforms.  However, despite their promise, their low-temperature behavior and magnetic properties remain largely unexplored and represent an underappreciated opportunity in quantum materials research.  With potential applications ranging from quantum computation to energy transfer, we believe that MOFs and particularly {\it magnetic} MOFs offer a vast and largely untapped frontier for transformative discoveries and high-impact quantum materials research.

\end{abstract}

\maketitle

%The discovery of high temperature superconductivity in 

The recent rate of discovery in solid-state physics has been remarkable.  From novel forms of superconductivity, recognition of the relevance of topology in materials, to new kinds of magnetism, such developments have changed our perspectives on the ways electrons can organize themselves in solids~\cite{Tokura2017}.  However, despite these successes, conventional approaches to realizing particular ground states in inorganic systems are still limited.  For instance, despite exhaustive search, there has been no definitive realization of a quantum spin liquid (QSL) state.  Hopes to find definitive signatures of them in many classes of frustrated lattices of interest~\cite{Broholm2020,Savary2016}~have not been realized. 
%{\color{red}Sara, I think you have added something about exfoliated samples, which got lost (?)}
And all known topological insulators are not -- in their native form -- actually insulators.  Conventional paradigms for materials assembly and tuning in inorganic systems are constrained as they are typically stable only over a small range of parameters: For example,  the ions that impart magnetism are also usually important for giving physical stability, which restricts their tunability. In this regard, we believe that  metal-organic frameworks (MOFs) are a promising platform  to find new quantum materials that circumvent many of these issues. In the areas of new materials, energy transfer, quantum computation, and quantum coherence, we believe there is tremendous potential for these materials to provide capabilities essential for next-generation technologies.  Although discussion on MOFs as quantum materials (``Quantum MOFs'') has already started, there is a vast unexplored landscape with literally thousands of  materials to be explored at low temperature~\cite{Huang2024}.

\begin{figure}[b]
	\includegraphics[width=0.9\textwidth]{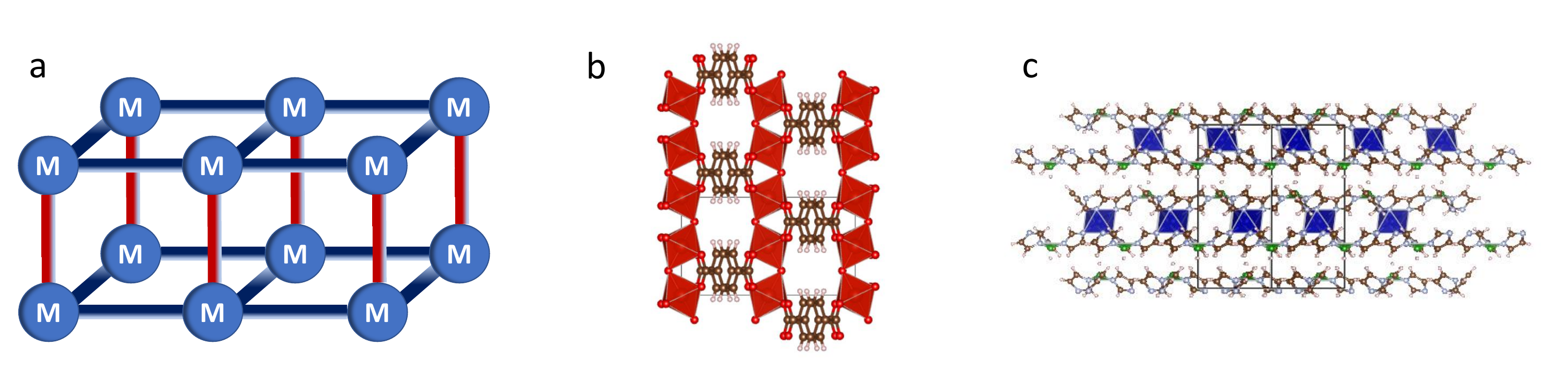} 
		\caption{(a) Schematic view of a magnetic metal-organic framework crystal structure: magnetic metal nodes (M) are connected into a crystal structure by organic linkers (blue and red bars). Linkers can be similar or different in- and out-of-plane. (b) An example of the crystal structure of MIL-47 containing V centers, where spin chains hosted by VO$_6$ octahedra are connected into crystollagraphic network by bdc (benzenedicarboxylic acid) molecules~\cite{Barthelet2002}.  (c) Crystal structure of  triangular lattice magnetic MOFs based on boron imidazolate frameworks (BIFs), Co- BIF and Ni-BIF. Triangular lattice Co-BIF shows antiferromagnetic interactions on the order of 1 K, and a spin-crossover-like effect in magnetic susceptibility due to thermal depopulation of excited crystal electric field levels. Magnetic properties of Ni-BIF suggest sizable ferromagnetic interactions~\cite{davis2024tunable}.  All structures visualized in VESTA. }
	\label{fig1}
\end{figure}

MOFs present an opportunity for a new paradigm for tunable quantum materials. They are a class of crystalline solids in which metal atoms or clusters interact through organic linkers, which connect them into a crystal structure (see Fig.~\ref{fig1})~\cite{Espallargas2018,Thorarinsdottir2020}.  MOFs are an  incredibly versatile materials family, which has been extensively investigated for other reasons including catalysis and hydrogen storage~\cite{qiu2020metal,suh2012hydrogen}. The general importance of MOFs was recently acknowledged by the Nobel Prize in Chemistry awarded in 2025 to Susumu Kitagawa, Richard Robson and Omar Yaghi for the discovery of these materials.

MOFs  are potentially transformative platforms for Hamiltonian engineering to realize quantum states of matter exhibiting entangled {\it quantum magnetism, superconductivity, and topology}.  Magnetism in MOFs typically appears as a result of the magnetic atoms or clusters as metal nodes in the structure. There are theoretical predictions of the rich physics to be realized in magnetic MOFs (mMOFs) including quantum spin liquids~\cite{Jacko2015},  Kitaev spin systems~\cite{Yamada2017,yamada2017crystalline}, superconductors~\cite{Jiang2021}, and exotic topological states~\cite{kambe2014redox,Jiang2021}.  However, despite these predictions and their vast promise, they are under appreciated by the quantum materials community.
 %\section{general MOF words}
%\section{spin moment tuning}
%\section{dimensionality}
While a fair number of magnetic MOFs have been synthesized (and even reviews written~\cite{Thorarinsdottir2020}), until recently there has only been modest effort in experimental investigation of these states at low temperatures.  The few examples include  %We know of only a few relevant, but disconnected works.  There was 
 reports of quantum spin liquid behavior in Cu-based 2D kagome lattice MOFs~\cite{Misumi2020,Pratt2024}, superconductivity at $T_c=0.25$~K   found in a different Cu-based kagome network ~\cite{Takenaka2021}, and a metal-azolate framework composed of tetrahedral units with five spins that was observed to order only at low temperature~\cite{nutakki2022frustration}.

Synthetic versatility of MOFs gives unique possibilities to pair different organic linkers with magnetic ions to control the geometry of the magnetic lattices.  The strength and nature of interactions between magnetic nodes can be tuned by chemical modifications of the linkers (e.g., length, geometry, ligand donor strength, electronic properties).  %Moreover, magnetic interactions are larger than one might expect considering the distance between magnetic nodes, which is due to strongly extended $\pi$ orbitals of the organic linkers. 
In some cases, they can exhibit magnetic order at temperatures as high as a few hundred Kelvin~\cite{Espallargas2018,Thorarinsdottir2020,Perlepe2020}. 
%\section{Linkers tuning}
%The strength and character of magnetic interactions in mMOFs can be varied over a wide range depending on the type of the linkers.  Some MOFs exhibit magnetic order as high as a few hundred degrees Kelvin~\cite{Espallargas2018,Thorarinsdottir2020,Perlepe2020}. 
In general, magnetic interactions are larger than one may expect considering the distance between magnetic nodes because the linkers, for instance benzencarboxylate~\cite{Thorarinsdottir2020} possess extended and strongly coupled $\pi$ orbitals.  These are the same aspects that endow these structures with efficient energy transfer, which is the key to their use in solar energy, transistors, batteries, supercapacitors, and catalysis~\cite{qiu2020metal,shi2019applications,mezenov2019metal}.  They are also very tunable:  carboxylate-containing frameworks offer the ability to add functionality, such as -OH, -SH, and -NH$_2$ groups, to tune the size of exchanges. These multitopic linkers can be shortened and elongated to adjust the distance between magnetic centers to tune the interactions.  An example of such chemical modification that led to the tuning of magnetic interactions by involving linkers of different sizes can be found in Ref.~\cite{Lim2017}. A few studies have shown that the interactions between magnetic metal nodes can be substantially enhanced by placing an unpaired spin on a linker~\cite{Perlepe2020,Zhang2022}. In Ref~\cite{Perlepe2020}  
Li$_{0.7}$[Cr(pyrazine)$_2$] Cl$_{0.7}$·0.25(THF) shows ferromagnetism, facilitated by the formation of a pyrazine radical with S = 1/2, at least up to 500~K, while the parent square lattice mMOFs  Cr(pyrazine)$_2$Cl$_2$ demonstrates ferrimagnetism below 55~K~\cite{pedersen2018formation}.   Additionally strain and pressure can be used to tune the exchange interactions.  Some MOFs are also photo-responsive, wherein their chemical and physical properties can be tuned {\it operando} by photoirradiation~\cite{Rice2019photophysics}. The sheer number of possible functionalities in MOFs is unparalleled in any other material class.

%%%
The variety of tunable properties includes dimensionality of the structures, with 1D, 2D, and 3D structures possible. 1D mMOFs have been demonstrated ~\cite{Thorarinsdottir2020}, but so far, fundamental understanding of magnetism in 1D mMOFs is limited.  Some examples are recent work on the demonstration of a Haldane chain behavior in [Ni($\mu$-4,4'- bipyridine)($\mu$-oxalate)]$_n$(NiBO)~\cite{Tin2023} and a study of a chiral S=1/2 chain [Cu(pym(H$_2$O)$_4$]SiF$_6$H$_2$O~\cite{Scatena2025}.  While hexagonal symmetry is not common in the world of inorganic crystals (with graphene and related materials being notable counter-examples),  MOFs can easily realize lattices with six-fold symmetry, such as kagome or hexagonal lattices, due to the fact that benzene rings and its derivatives are popular building blocks of MOFs. These lattices are important in the field of frustrated magnetism, as structures to realize  quantum spin liquid behavior.  Hexagonal lattices may host Kitaev spin liquids~\cite{Takagi2019} in systems with sufficiently large spin-orbit interactions.  Despite existing predictions about Kitaev spin liquids in mMOFs~\cite{yamada2017crystalline} and some  progress in identifying possible spin-liquid kagome lattices ~\cite{Misumi2020,Pratt2024}, detailed experimental studies and material synthesis attempts are still to come.

Different physics may be accessed depending on the spin character of the magnetic ion (e.g. $S = 1/2$, 1, 3/2 etc.), which tunes the degree of on-site quantum fluctuations. Magnetic nodes can be easily chemically exchanged  while preserving crystal structure and linkers, allowing to study the effect of different spin character~\cite{Davis2024a}.  Ideal experiments can imagine being done were one tunes a 1D magnetic chain by only varying the spin magnitude  (S=1/2,1, 3/2) through substitution or via external stimuli such as light interactions, while keeping all other parameters fixed.
%An example of such MOFs is the family of MIL-47(M) materials (M=metal ion), where the crystal structure  consists of 1D metal oxide chains connected into in 3D crystal structures by organic linkers.

These tantalizing examples engender great optimism regarding potential discovery of new physics. In addition to magnetic states like quantum antiferromagnets, topological Kitaev spin liquids, and kagome spin liquids, phases to be realized in these materials includes topological superconductors, topological insulators, and other exotic quantum states of matter.  Kagome lattices are of high interest not only for magnetic MOFs, but also as a host of flat bands~\cite{Fuchs2020}. In particular, superconductivity as a consequence of electronic correlations enhanced in the flat bands is predicted for Kagome MOFs~\cite{Ohlrich2025}. Flat bands as such may have been observed by STM ~\cite{Hu2023}.  The vast parameter space, though daunting, is an incredibly promising topic for new materials exploration with a high likelihood of transformative scientific impact.

% There is a zoo of literally thousands of  materials to be explored at low temperature.

There is lots to be done, particularly with regards to refining single crystal growth and spectroscopic studies.   With regards to sample morphology, MOFs of interest can be single crystals or made into large area films~\cite{kim2019uniform,kim2020centimeter}, including single layer purely 2D samples where interaction with a substrate can play a role for the modifying properties  \cite{Field2022}.  
 Encouragingly, the synthetic procedures to generate reliable and scalable MOFs have advanced drastically beyond typical solvothermal methods. Microwave-assisted and mechano-chemical techniques, for instance, have enabled more atom-efficient and energy-saving syntheses of MOFs, where the solvent is either eliminated or significantly reduced in volume and multi-day heated syntheses can be completed in minutes. However, procedures that improve yields and rates often lead to bulk powders.   The challenge for obtaining large single crystals of MOFs is to achieve conditions that decrease the rates of nucleation and growth; fast nucleation and growth rates lead to the accumulation of crystal defects and twinning~\cite{Vinogradov2021}.  Strategies to obtain high-quality single crystals are actively being explored, with tuning of chemical and physical parameters such as solvent/precursor identity, reactant concentrations, temperature, and reaction volume.  For instance, ``modulators," which are typically monotopic ligands, can compete with the multitopic linkers and decrease the nucleation rate and limit growth in different directions~\cite{Bigdeli2023}. Substrate-supported growth of MOFs, such as chemical vapor deposition (CVD) or molecular epitaxy, can also be leveraged to synthesize large-area thin films~\cite{Bradshaw2012}. In CVD, relatively low boiling point precursors are vaporized and flowed over a substrate to start the nucleation process~\cite{Han2020, Ameloot2016}. By carefully controlling the temperature and flow, the diffusion of the reactants is precisely tuned.  Molecular epitaxial growth of MOFs involves the layer-by-layer introduction of chemical precursors, either through the liquid phase (e.g., dipping, spraying) or the gas phase (akin to atomic or molecular layer deposition)~\cite{Choe2024,woll2012}. 

%%{\color{red} Alternatively, CVD growth of thin films of MOFs provides a way to produce large area high quality samples, having the advantage of a control over the speed of growth by the vapor density (or whatever is the parameter)  ~\cite{Han2020} }
 %%Molecular epitaxial growth of MOFs has gained significant interest owing to the ability--need to finish-st }
%%{\color{blue} More references on CVD and crystal growth, the reviewer wanted to have MBE growth, I am not sure anyone has done that.- MLD (molecular layer deposition, is it a molecular alternative for MBE?)~\cite{Choe2024}}

 Most of the experimental information about conducting and magnetic MOFs has come from traditional bulk measurements such as resistivity and magnetic susceptibility. In order to probe the properties of MOFs as quantum materials, spectroscopic information is going to be essential. Optical spectroscopies have been an effective way to probe the molecular and lattice vibrations of MOFs in the past~\cite{Hadjiivanov2020,Sunil2023,Dziobek2025}. These can be extended to probe relevant low-energy electronic and magnetic excitations, as well as excitations in topological bands, following the techniques developed for studies of inorganic correlated electron systems: Infrared, THz, and Raman scattering spectroscopies are successfully used to study magnetic properties, changes of electronic structure on metal-insulator transitions and superconducting gap~\cite{Devereaux2007,Basov2011}, electronic bands and their topology~\cite{Drichko2015,armitage2019matter}. STM spectroscopy has also emerged recently as a successful spectroscopic technique to probe electronic structure in MOFs~\cite{Lowe2024,Hu2023}. In frustrated magnetism research, neutron scattering has been accepted as the gold standard to prove exotic magnetic states~\cite{Bramwell2011}. While it cannot be used as a high throughput method for probing magnetism in these materials due to the presence of hydrogen in the structure, which is a high neutron scatterer, the recent progress in the MOFs crystal growth and a possibility of growing deuterated samples can make measurements of significant materials possible~\cite{Zheludev2001}.

With the support of the Army Research Office, we recently held a workshop entitled, ``Metal-Organic-Frameworks as Quantum Materials," at Johns Hopkins University on May 7-9, 2025, with the goal of bringing a diverse community of researchers to give their unique perspectives on mMOFs as quantum materials \footnote{\url{https://sites.google.com/view/mofasqm/home?authuser=0}} \footnote{5/7 Morning: \url{https://drive.google.com/file/d/10lkgX8A41eYVmYtLN3lF2FELeoJYEXlH/view?usp=sharing}\\
5/7 Afternoon: \url{https://drive.google.com/file/d/1JnUS7lzAIpsIxgse0Viqvey8Q3kiok0o/view?usp=sharing} \\
5/8 Morning: \url{https://drive.google.com/file/d/1vcwOVociZNy2l0HA3EXTp8wGFAk9vbMA/view?usp=sharing} \\
5/8 Afternoon: \url{https://drive.google.com/file/d/1QtPkxCxDkyAqVvHC5u7y_u-1BT3Nt1Xz/view?usp=sharing}\\
 5/9 Morning: \url{https://drive.google.com/file/d/1iTxwmOALGdjbgm_Xf1SniHXRfNjdAsVt/view?usp=sharing} }
\footnote{Slides from talks can be found here \url{https://drive.google.com/drive/folders/177NE8L08PacMEjNSm8nyO-O2tsiHnRKP?usp=sharing}.}.  Over 2.5 days, synthetic chemists, condensed matter spectroscopists, chemical engineers, and theoretical physicists (among others) shared current results, brainstormed, and gave their thoughts on the future of this emerging field.  A number of speakers (Pedersen, Thoi, Oppenheim, Huang, Morris) emphasized the wide synthetic landscape of MOFs--from atomic alloying to ``defect" engineering--for applications in magnetism, energy storage, charge conduction, photocatalysis, and energy transfer.~\cite{Afrin2024,wang2023dominant,streater2023wavelength}  For instance, Pedersen discussed his group's work characterizing magnetism in a family of frameworks based on metal pyrazine and the ability to ``alloy" them for controlling magnetic interactions. \cite{Perlepe2020,Aribot2025} Oppenheim discussed synthetic strategies to tune and control electron transport in 3-dimensional MOFs.~\cite{Zhang2025,Xie2018} Relatedly, McQueen and Bukowski discussed the unique connections between structure and properties in MOFs.~\cite{Berry2022,Yang2023}

The talks of other speakers (Manas-Valero, Kempa, Shatruk, Feng) pointed out the tremendous possibilities for new functionality in this material family.~\cite{Mazarakioti2025}  Kempa discussed his group's work using MOFs as templates for potential shaping in an adjacent 2D material layer.~\cite{Kingsbury2025} Shatruk pointed out the that lanthanide nodes in MOFs may work used as an array of clock-transition qubits that may be protected from decoherence caused by magnetic noise.~\cite{Gakiya2025}  Feng discussed opportunities with 2D conjugated MOFs for applications in the form of ``MOFtronics".~\cite{Fu2025,Zhang2023,wang2024cu3bht} And the physicists (Zapf, Musfeldt, Jaubert, Drichko) discussed efforts to predict and find new and interesting phenomena in these compounds.~\cite{davis2024tunable}  Musfeldt detailed her group's work on understanding the details of the phase diagram on a multi-ferroic MOF [(CH$_3$)$_2$NH$_2$]Mn(HCOO)$_3$~\cite{Clune2020}, while Zapf demonstrated the potential of the same Mn MOFs for magnetoelectric applications.~\cite{jain2016switchable} Jaubert  made the point that MOFs naturally host structures that may be seen as a 3D embedding of a pentachoron, which is a 4D equivalent of the tetrahedron.  He described a spin liquid state that is described by a 3D slab of a 4D Coulomb gauge field and proposes that metal-azolate frameworks [Mn(ta)$_2$] may realize it~\cite{nutakki2022frustration,nutakki2023classical}. 

Despite the tremendous promise, challenges to working in this field abound.   These include the issues associated with working with delicate samples, the dearth of large single crystals, and the difficulties of working with standard condensed matter physics tools for connecting structure and magnetism in samples with lots of hydrogen (e.g. neutron scattering).  They also include the social challenges of bringing together diverse groups of scientists who may have essential skills or expertise, but who may have different primary interests or speak with very different terminology for the same phenomena.  Of course, ideally these differences are a strength.  We anticipate the ongoing exchanges between scientists working from different perspectives will make this a dynamic -- and possibly revolutionary -- field in the years to come.

\section{Funding declaration}

The workshop was supported by a grant from DEVCOM Army Research Laboratory- Army Research Office (ARO)’s Solid State Physics (SSP) Program.  ND, ST, and NPA's work on this topic is supported by the ARO via grant W911NF2510033 ``Metal-organic frameworks (MOFs) as designer quantum materials".

\section{Author contributions}
 All authors contributed equally to the writing the text of the manuscript

\section{Competing interests}
The authors declare no competing interests.

\bibliographystyle{apsrev}
\bibliography{MOFreferences}

\end{document}